\documentclass[a4paper]{article}

\usepackage{INTERSPEECH2021}
\usepackage{color}

\title{Automated detection of foreground speech with wearable sensing in everyday home environments: A transfer learning approach}
\name{Dawei Liang, Zifan Xu, Yinuo Chen, Rebecca Adaimi, David Harwath, Edison Thomaz}
\address{
  University of Texas at Austin}
\email{dawei.liang,zfxu,ynchen2829,rebecca.adaimi,ethomaz@utexas.edu, harwath@cs.utexas.edu}

\begin{document}

\maketitle
\begin{abstract}
Acoustic sensing has proved effective as a foundation for numerous applications in health and human behavior analysis. In this work, we focus on the problem of detecting in-person social interactions in naturalistic settings from audio captured by a smartwatch. As a first step towards detecting social interactions, it is critical to distinguish the speech of the individual wearing the watch from all other sounds nearby, such as speech from other individuals and ambient sounds. This is very challenging in realistic settings, where interactions take place spontaneously and supervised models cannot be trained \textit{apriori} to recognize the full complexity of dynamic social environments. In this paper, we introduce a transfer learning-based approach to detect foreground speech of users wearing a smartwatch. A highlight of the method is that it does not depend on the collection of voice samples to build user-specific models. Instead, the approach is based on knowledge transfer from general-purpose speaker representations derived from public datasets. Our experiments demonstrate that our approach performs comparably to a fully supervised model, with 80\% F1 score. To evaluate the method, we collected a dataset of 31 hours of smartwatch-recorded audio in 18 homes with a total of 39 participants performing various semi-controlled tasks.

\end{abstract}

\vspace{-3pt}
\section{Introduction}
\vspace{-3pt}

Researchers have recently shown that off-the-shelf smartwatches can serve as an effective platform to recognize various forms of human behaviors and activities of daily living \cite{becker2019gestear, thomaz2015inferring}. In our research, we are particularly interested in exploring whether the acoustic sensing capability of smartwatches can be used to detect and characterize face-to-face social interactions. Quantifying social interactions is important in numerous applications, ranging from understanding workplace dynamics \cite{nadarajan2019speaker} to identifying vocal biomarkers of cognitive impairment \cite{konig2015automatic}.

A critical step towards detecting in-person social interactions of an individual (i.e., an individual wearing the smartwatch) is to distinguish their speech from other sounds. In wearable sensing, speech produced by the person wearing the watch is usually referred to as \textit{foreground} speech while all other forms of noise such as the speech of others, natural sounds, man-made sounds, music, and silence are often referred to as the \textit{background} \cite{wrigley2004speech}. Hence, the problem of foreground detection in wearable sensing can be formulated as binary classification of foreground speech versus background sounds. 



As previously noted by Nadarajan et al. \cite{nadarajan2019speaker}, an intuitive approach to detecting foreground is to employ a speaker verification system with a voice activity detector. Such a system would be able to map speech to one or more speakers \cite{Bredin2020}. Similar efforts include estimating a binary mask of target speech \cite{wang2018supervised}. However, these techniques would have to assume that the user or channel is a known \textit{apriori}. This is often not possible in non-deterministic real-world settings, when conversations and interactions cannot be anticipated ahead of time. Instead, the detection of foreground speech must take place in a speaker-agnostic manner \cite{nadarajan2019speaker}. Another direction in foreground detection with wearable sensing is to build a supervised model that learns to classify foreground speech versus background sounds \cite{hebbar2021deep, little2020deep}. This technique, however, requires annotated training sets collected in real-world environments. In practice, due to the difficulty of compiling a dataset that is representative of all possible situations and sounds, this technique fails to generalize \cite{hebbar2021deep}. 

In this paper, we explore a different formulation to this challenging problem. We propose and evaluate a foreground speech detection method that does not require training a model on specific foreground datasets. Our contributions are: 

\begin{itemize}

    \item A speaker-agnostic foreground detection method based on clusters of acoustic embeddings; the embeddings are derived from deep features extracted from a public speaker dataset, and can capture foreground speech characteristics without being explicitly trained with audio of the foreground speakers and others, i.e., without supervision. 
    
    \item An evaluation of the proposed method with social interaction audio collected  by 39 participants in their own homes. We obtained an F1 score of 80\%, demonstrating that the method performs comparably to one based on a fully supervised model \cite{nadarajan2019speaker}.
    
    \item The audio dataset\footnote{https://github.com/Human-Signals-Lab/Foreground-speech} we collected in this research, which contains 32 hours of foreground and background sound classes recorded by smartwatches in real homes in a semi-controlled manner.
    
\end{itemize}

\vspace{-3pt}
\section{Related work}
\vspace{-3pt}


Early attempts at inferring target speech instances without accessing the user voice fingerprints include comparison of sound energy obtained from multiple devices \cite{wyatt2007privacy}, or by using zero-frequency filtered signals \cite{deepak2012foreground}. However, such approaches were generally deployed for the lab environment with well-defined forms of noise, and the features do not capture foreground speech related characteristics. More recent approaches towards foreground detection in wearable sensing mostly focused on developing a foreground detector directly with customized data collected from the target environment \cite{hebbar2021deep, little2020deep}, but this inevitably requires a large effort of study deployment and data annotation which is hard to scale. 

Our work explores knowledge transfer from speaker recognition to foreground speech detection. Transfer learning from a source domain to a target domain has been shown to be useful for the target task when knowledge obtained from the source domain is not present in the target \cite{tan2018survey}. Such knowledge transfer can be across datasets \cite{liang2019audio}, tasks \cite{dissanayake2020speech, liang2021transferring} or even learning modalities \cite{aytar2016soundnet}. For foreground detection, Nadarajan \textit{et al}. \cite{nadarajan2019speaker} proposed an approach by training a foreground detector on a public meeting dataset and generalizing it to customized meeting scenarios in a hospital with an audio recorder worn on the participants' chest. As discussed by later work \cite{hebbar2021deep}, however, such approach does not generalize well for wearable sensing in daily living scenarios. 



\vspace{-3pt}
\section{Dataset}

\vspace{-3pt}

This section describes the audio dataset we collected in order to evaluate our approach. The dataset captures common foreground and background sound classes collected by a smartwatch in realistic home settings.



\subsection{Data collection protocol and procedures}

Our data collection was based on a semi-controlled study. This IRB-approved study took place in 18 distinct homes, each containing at least two participants. For each group, one of the participants was asked to wear a Fossil smartwatch for continuous audio recording, and this participant was referred to as the wearer of the study. The other participant(s) in the home engaged in a set of social activities with the wearer following a pre-defined study script. Table \ref{session} lists the activity sessions.


\begin{table}[t]%
\caption{The studied interaction sessions in homes.}
\vspace{-6pt}
\begin{center}
\begin{tabular}{ll}
  \toprule
  \small \textbf{Session} & \small \textbf{Description} \\
  \midrule
    Telephone Call  & The wearer placed a telephone call \\
    & with another remote researcher. \\
    
    Watch TV & The wearer watched content on a TV/laptop \\
    & with the sound on for 10min. \\
    
    Chat Indoors  & All participants played the NASA \\
    & decision-making game \cite{nasa2020} around a table. \\
    
    Chat Outdoors  & All participants walked outdoor for at \\
    & least 5min and chatted on any topics. \\
    
    Meal  & All participants (including household \\
    & members) cooked and ate together for dinner.\\

  \bottomrule
\end{tabular}
\vspace{-12pt}
\label{session}
\end{center}

\end{table}%

The study period was from 3:30pm to 8pm in a day, where there was a 30-minute gap between each of the sessions. To maintain the naturalistic nature of the interaction process, the watch was left recording continuously throughout the entire study period, and we did not specify a set time for the conversations to end. However, the wearer was asked to note down the approximate end time of each session. The acoustic environments were left as usual, where sounds of home appliances and other non-participating household members were allowed to be included anytime throughout the study. 

On the day of the study, the study materials including the smartwatch, reading materials and study script were delivered by a researcher. Right before the study began, the researcher connected with the participants via Zoom to introduce the study and procedure. The researcher then logged out and collected the materials after the entire study completed. An audio recording app was pre-installed on the watch to enable continuous audio recording. Our preliminary test showed that the app can continuously record audio for up to 8 hours, so the battery was not a problem. The data was saved on the watch in PCM format with one 16-bit channel at a 16 kHz sampling rate.

In total, we collected data from 18 homes (groups). Groups 8, 9 and 10 consist of three participants including the wearer. The remaining groups consist of two, resulting in a total of 39 participants. The total number of household members per group varies from two to five. The participants age from 15 to 59, with various occupations. Besides, 19/39 of the participants were male, and 15/18 of the wearers were right-handed. All study participants were fluent in English or native English speakers.

\vspace{-2pt}
\subsection{Annotation}
\vspace{-2pt}

We annotated the collected audio on a server after each group of collection by listening back to the audio clips. The gap of the sessions was not used. We annotated the audio with a temporal  window of one second with no overlap. The fine-grained labels included \textit{wearer speech}, \textit{non-wearer speech}, \textit{ telephone voice}, \textit{television}, \textit{mixed speech}, \textit{baby sounds}, \textit{non-vocal noise} and \textit{ambiguous sounds}. Specifically, class \textit{wearer speech} indicated the case if the speech turn within an instance was exclusively held by the wearer, and \textit{mixed speech} indicated an overlap between the wearer and other speech. \textit{Non-wearer speech} was used when the speech turn was only held by non-wearer physical participants. Other observed vocal sounds included \textit{telephone voice}, \textit{television}, or \textit{baby sounds}. Instances of silence or non-vocal background noise were labeled as \textit{non-vocal noise}. \textit{Ambiguous sounds} was used if the annotators were not confident about the sound type. Following the previous definition, we categorized classes \textit{wearer speech} and \textit{mixed speech} both as the foreground speech type, whereas instances of all other vocal and non-vocal classes were counted as the background type. The annotation reached a mean of 0.907 Cohen's kappa inter-rater reliability for the group of annotators, based on evaluation of a randomly selected interaction session. The kappa value indicates a good agreement of annotation \cite{landis1977measurement}. 

\begin{table}[t]%
\caption{Instance size distribution of sound classes. }
\vspace{-6pt}
\begin{center}
\begin{tabular}{lll}
  \toprule
  \small \textbf{Labels} & \small \textbf{\# of instance} & \small \textbf{Percent} \\
  \midrule
    Non-vocal noise & 36,801 & 38\%\\
    Television & 21,131 & 22\%\\
    Wearer speech & 19,461 & 20\%\\
    Non-wearer speech & 13,041 & 13\%\\
    Mixed (wearer\&others) speech & 2,600 & 3\%\\
    Telephone voice & 1,903 & 2\%\\
    Ambiguous sounds & 1,146 & 1\%\\
    Baby sounds & 688 & 1\%\\

  \bottomrule
\end{tabular}
\vspace{-12pt}
\label{instance}
\end{center}

\end{table}%

We obtained a total of 31 hours (111,423 1-second instances) of audio in our dataset. We categorized 23.5\% of the total instances as the foreground speech instances. Ambiguous instances accounted for only 1\% of the total. Table \ref{instance} shows the temporal size distribution of the collected sound classes based on the fine-grained annotation. The imbalance in temporal distribution of the sound classes indicates the nature of people's unconstrained social behaviors. Our released dataset contains the Fast-Fourier Transform (FFT) features and transfer learning embeddings extracted from the audio. Details of the features will be described below.


\vspace{-3pt}
\section{Methodology}
\vspace{-3pt}

\begin{figure}[t]
\centering
    {\includegraphics[width=\linewidth]{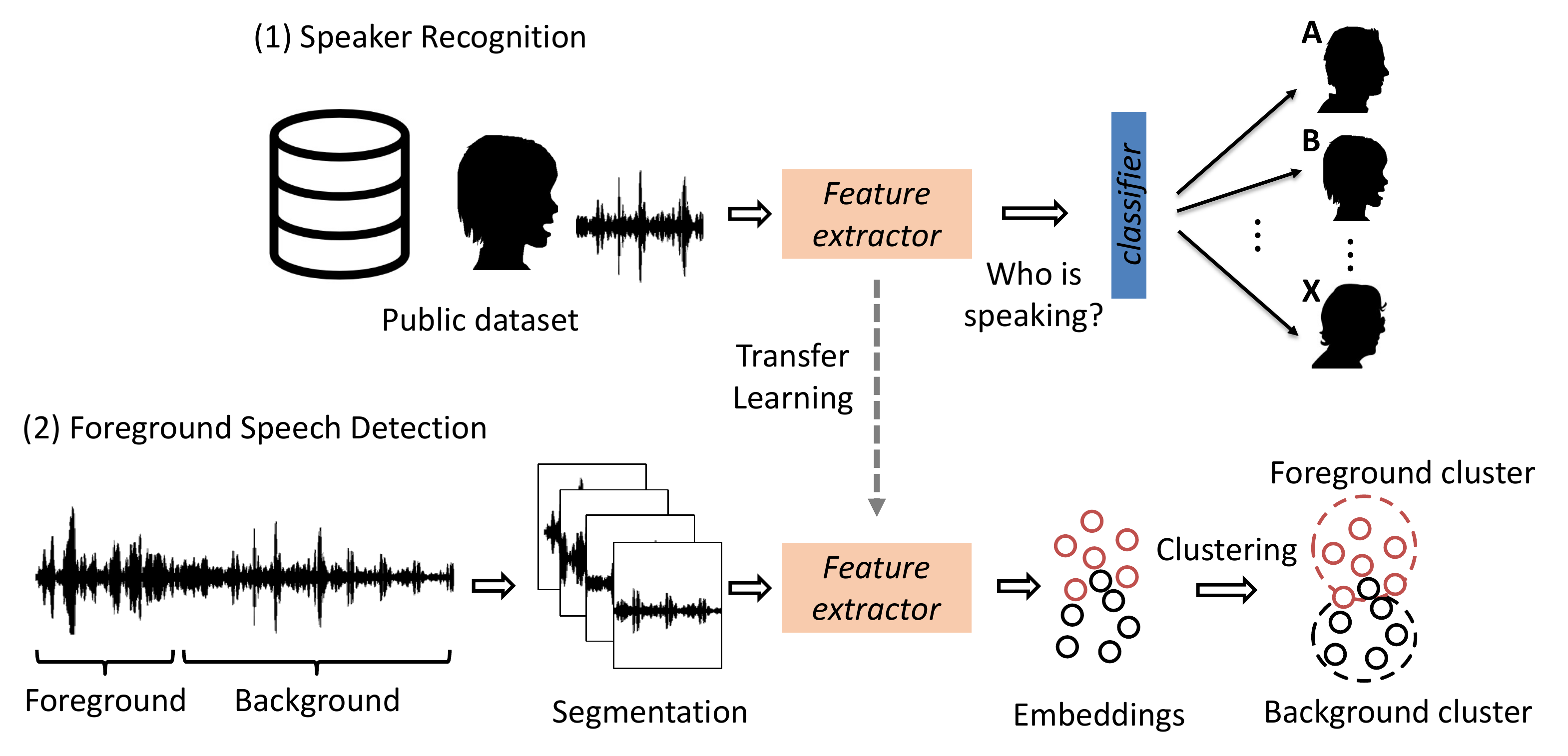}}
  \caption{Overall pipeline of our study. The source task is general-purpose speaker recognition based on a public speaker corpus, and the target task is foreground detection towards our smartwatch audio recordings.}
  \vspace{-3pt}
  \label{pipeline}
\end{figure}

Figure \ref{pipeline} shows the overall pipeline of our study. It consists of two major steps, (1) development of a speaker model with a public speaker dataset which serves as a feature extractor for foreground detection on our collected data, and (2) inference of foreground instances using a clustering-based approach. The first step pre-trains a neural network on a public dataset to classify given speaker classes of the dataset. This is the process for the network to learn specific voice representations from the input audio. Once trained, the network is fixed and generalized to our customized foreground detection task. The second step takes as inputs our smartwatch audio consisting of the foreground and background instances and performs clustering on the computed embeddings. Although the network was not trained with any foreground and background contexts, our experiments show that it is capable of deriving foreground representations. The foreground and background clusters can be identified based on the overall distributions of the cluster samples.

Our motivation for leveraging embeddings in this work is that for instances collected from a foreground speaker, or even multiple foreground speakers wearing the same watch, the resulting voice characteristics are generally consistent in a sense that they are mostly captured in close proximity to the watch microphone. Such in-close-proximity voice can trigger the feature extractor to generate embedding features representing similar voice classes from its pre-trained pool. For the background instances that are mostly captured in the far field, including more disrupted background speech from other speakers, non-vocal noise or silence, the corresponding sound characteristics are generally not covered in the pre-trained pool, resulting in embeddings of a null-class type. Hence, such differences of embedding types may be significant enough to be captured by a clustering process. Admittedly, it can be more error-prone when the foreground and background speech is both captured close to the smartwatch. However, our experimental results show that the proposed approach can yield a comparable performance to a supervised model.

\subsection{Feature Extractor}

We studied two strategies to derive our feature extractor in the source domain. The first strategy was based on the Pyannote speaker diarization toolkit \cite{Bredin2020}. Specifically, we used its speaker model \cite{Bredin2017} pre-trained on the public VoxCeleb 1 \& 2 \cite{nagrani2017voxceleb, chung2018voxceleb2} datasets. Both datasets consist of audio utterances of over 1K celebrities from public YouTube videos of varying lengths. The second strategy was based on a shallow neural network feature extractor \cite{lukic2017learning} that we trained on a random subset (65 males and 35 females) of the public TIMIT \cite{garofolo1993timit} clean speech corpus. For the second strategy, specifically, we augmented the clean speech utterances with common household noise for better model generalization \cite{ko2015audio}. The noise clips were collected by a researcher using a smartphone placed in the home, including \textit{balcony, chopping, dining room, frying food, having lunch, preparing meals, strolling outdoor, watching TV, washing (with water), white noise}. The clips were 30 seconds each, except for \textit{having lunch, preparing meals} and \textit{watching TV} which had 60 seconds each. We then copied each TIMIT utterance for multiple times and overlaid each copy with a noise clip at one of the signal-to-noise levels of 3 db, 10 db, and 20 db, respectively.

The audio utterances were truncated to one second for both strategies to match the input requirement of our foreground detection task. The pre-trained model of strategy 1 generates 1D embeddings of size 512 per second, which we refer to as \textit{emb1}. The model of strategy 2 takes as inputs (128$\times$100) spectrogram features from the raw audio for each second. We applied an 80\%-20\% training-test split on each of the TIMIT subjects to train the network until the validation accuracy stopped improving. In the foreground detection experiments, we obtained 1D embeddings from its first fully-connected layer which have a size of 1,000 per second. We refer to embeddings generated by the second strategy as \textit{emb2}.

\subsection{Clustering Embeddings}

As previously indicated, our assumption for foreground speech detection is that we cannot assume a supervised model to be developed on a customized foreground audio dataset, and no apriori data from users is available. In this step, we directly applied the feature extractors developed above to obtain embeddings from the collected smartwatch audio. Next, we explored if a clustering process can be used to identify the  foreground and background embedding instances. In our experiment, we used cosine similarity as the distance measure of the embeddings. We tested K-Medoids and spectral clustering based on Python Scikit-learn and Scikit-learn-extra packages \cite{scikit-learn, scikitlearnextra} since they are well suited for cosine similarity measure. The number of output clusters was pre-defined, which was two for foreground detection. Once the output clusters were formulated, we compared the average similarity from centroid for each output cluster to identify its class label, which is calculated by (\ref{equ1}):

\begin{equation}
 \label{equ1}   
\overline{S} = \frac{\sum S(\textbf{\textit{x}}, \overline{\textbf{\textit{x}}})}{n}  \end{equation}

 Where $\overline{S}$ denotes the average cosine similarity from centroid. $S(\textbf{\textit{x}}, \overline{\textbf{\textit{x}}})$ is the cosine similarity between a cluster sample and the cluster centroid, $\overline{\textbf{\textit{x}}}$, and n is the total number of cluster samples. The average similarity from centroid measures the variability of the cluster. Based on our preliminary studies, the resulting $\overline{S}$ for a background cluster was always bigger than the $\overline{S}$ for the corresponding foreground cluster of a user. This is possibly because the null-type-like background embeddings are generally less distinguishable than the foreground embeddings which are abstracted from the pre-trained voice pool, resulting in an overall higher within-cluster similarity. Hence, we applied this same criterion to label the output clusters.

\vspace{-3pt}
\section{Experiments and Results}
\vspace{-3pt}

\begin{figure}[t]
\centering
    {\includegraphics[width=\linewidth]{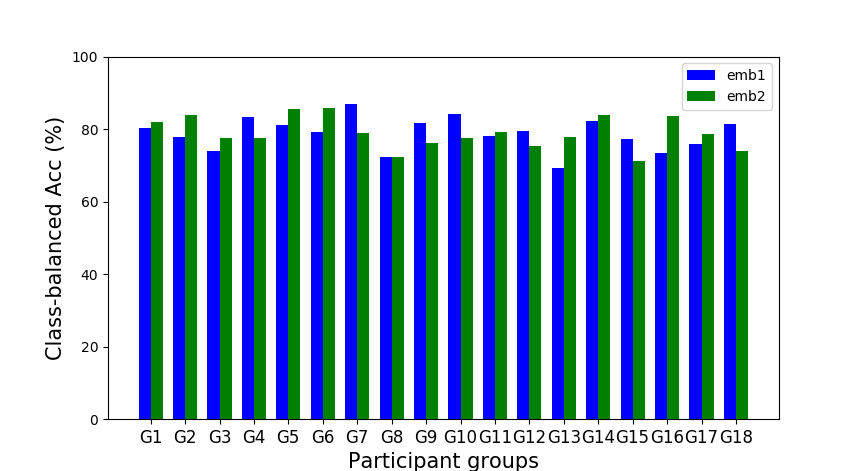}}
  \caption{Group-wise performance based on different types of embeddings, using spectral clustering.}
  \vspace{-6pt}
  \label{group}
\end{figure}

We first studied the overall foreground detection performance using the proposed method. We reported the class-balanced accuracy and macro F1 score to account for class imbalance. The class-balanced accuracy/macro F1 is calculated by averaging the recall/F1 values of the two target classes without weighting. Table \ref{unsupervised} shows the overall results. First of all, we can see that the overall results obtained by the proposed unsupervised approach are generally reasonable, around 80\% of accuracy and F1 scores. The results are consistent for both emb1 and emb2. Secondly, spectral clustering is the most robust method. The best accuracy was obtained by emb2 + spectral clustering, while the best F1 score was obtained by emb1 + spectral clustering. Thirdly, foreground detection with emb2 is more reliable. This is expected since the feature extractor of emb2 was pre-trained with more realistic noise conditions derived from everyday scenarios. The benefit of our customized data augmentation overwhelms the lack of speaker scale in the source domain. Figure \ref{group} further visualizes the group-wise accuracy for emb1 and emb2 based on spectral clustering. The results indicate that there is no significant difference between the two feature extraction strategies (p $<$ 0.05) with spectral clustering. 

\begin{table}[t]%
\caption{Overall foreground detection performance based on different embedding types and clustering methods.}
\vspace{-6pt}
\begin{center}
\begin{tabular}{lll}
  \toprule
  \small \textbf{Methods} & \small \textbf{Class-balanced acc} & \small \textbf{Macro F1}\\
  \midrule
    emb1+K-Medoids & 69.3\% & 64.2\% \\
    emb1+Spectral  & 78.8\% & \textbf{80.4\%} \\
    emb2+K-Medoids & 78.5\% & 78.1\% \\
    emb2+Spectral  & \textbf{78.9\%} & 79.0\% \\

  \bottomrule
\end{tabular}
\vspace{-12pt}
\label{unsupervised}

\end{center}

\end{table}%

\subsection{Supervised benchmark}

\begin{table}[t]%
\caption{Supervised leave-one-group-out evaluation for our smartwatch dataset, model based on VGG-slimmer \cite{nadarajan2019speaker}.}
\vspace{-6pt}
\begin{center}
\begin{tabular}{lll}
  \toprule
  \small \textbf{Input type} & \small \textbf{Class-balanced acc} & \small \textbf{Macro F1} \\
  \midrule
   emb1 &  76.3\% & 77.2\% \\
   emb2 &  78.6\% & 79.7\% \\
   FFT &  80.4\% & 81.5\% \\
  \bottomrule
\end{tabular}
\vspace{-12pt}
\label{supervised}

\end{center}

\end{table}%

To better compare the performance of our method, we ran a supervised evaluation following a leave-one-group-out (LOGO) scheme, where all but one group of participants were used to derive checkpoint models, and the remaining group was used for evaluation. The global result was then obtained by averaging the best evaluation results of individual groups. These results also serve as a supervised benchmark for our dataset.

We applied the VGG-slimmer \cite{nadarajan2019speaker} architecture and its training setup for the LOGO evaluation. We tested the model with three types of features extracted from our data - FFT features, emb1, and emb2. The FFT features were extracted based on a window size of 50ms with the same hop and 64 output bins. This results in FFT features of shape (64$\times$20) per second. We used the 'same' padding mode in Keras \cite{chollet2015keras} and removed max pooling for the last convolutional layer of the FFT model to avoid feature size mismatch. For emb1 and emb2, we switched the 2D layers as 1D to roughly match the model design. The kernel/stride was 3/1 for the 1D convolutional layers and 2/2 for 1D max pooling. We did not apply post processing on the model outputs. Table \ref{supervised} shows the supervised results. We can see that the results obtained by our proposed unsupervised method are comparable to the LOGO benchmark, yet those results are obtained without any supervision on the user dataset.

\subsection{Effectiveness of speaker embeddings}

To better understand the mechanism of foreground detection with the transferred speaker embeddings, we conducted extra two experiments. The first was to randomly pool two or three groups of study participants’ data together for foreground detection. This is an important aspect to investigate if the transferred embeddings truly capture the foreground speech characteristics rather than just voice patterns generated by the same primary speaker. In this experiment, speech from all pooled wearers was annotated as the same foreground class, whereas all other sounds were categorized as the same background class. We then extracted embeddings of the best accuracy (emb2) from the audio and applied spectral clustering for foreground detection. The detection performance was generally not disrupted - 78.3\%/77.8\% acc/F1 for pooling of two groups and 77.9\%/77.9\% acc/F1 for pooling of three groups, indicating that the 
transferred embeddings truly capture the foreground speech characteristics from the smartwatch audio data. 

We then studied if general-purpose acoustic embeddings can also be used for foreground detection in the same way. To this end, we replaced the proposed speaker feature extractors with YAMNet, a general-purpose acoustic embedding extractor pre-trained with AudioSet \cite{gemmeke2017audio}. However, by testing with spectral clustering, we found that the clustering process was not able to separate the foreground and background instances (acc/F1: 49.9\%/44.6\%), indicating that our motivation to pre-train the feature extractors with the source speaker knowledge is necessary for the transfer learning process.





\vspace{-3pt}
\section{Conclusion}
\vspace{-3pt}

This paper presented an approach for speaker-agnostic foreground speech detection towards smartwatch audio recordings and the associated smartwatch audio dataset collected from 39 participants in their homes. The dataset contains foreground and background audio features of common in-person social event types in the home. Based on our experiments, we showed that the proposed unsupervised approach is comparable to a fully-supervised model for foreground speech detection without accessing any user or channel apriori. We further discussed in detail the effects of embedding extraction strategies and their overall/group-wise performance.


\bibliographystyle{IEEEtran}

\bibliography{mybib}


\end{document}